\documentclass{article}
\usepackage{spconf,amsmath,graphicx,multirow,subcaption,caption,array}
\usepackage[breaklinks=true,letterpaper=true,colorlinks,bookmarks=false]{hyperref}
\usepackage{color}
\newcommand{\modelname}{SALT}

\newcommand{\et}{\emph{et al.}}

\newcolumntype{"}{@{\hskip\tabcolsep\vrule width 1pt\hskip\tabcolsep}}
\newcommand\blfootnote[1]{%
  \begingroup
  \renewcommand\thefootnote{}\footnote{#1}%
  \addtocounter{footnote}{-1}%
  \endgroup
}

\title{\modelname{}: Distinguishable Speaker Anonymization Through \\Latent Space Transformation}
\name{Yuanjun Lv$^{1,2}$, Jixun Yao$^{1,2}$, Peikun Chen$^1$, Hongbin Zhou$^2$, Heng Lu$^2$, Lei Xie$^1$\sthanks{Corresponding author.}}
\address{
$^1$Audio, Speech and Language Processing Group (ASLP@NPU), School of Computer Science, \\ Northwestern Polytechnical University, Xi'an, China\\
$^2$Ximalaya Inc., China\\
Xizhang (Shanghai) Network Technology Co., Ltd. \\
}

\begin{document}
\copyrightnotice{979-8-3503-0689-7/23/\$31.00~\copyright2023 IEEE}
%\ninept
%
\maketitle
\begin{abstract}

Speaker anonymization aims to conceal a speaker's identity without degrading speech quality and intelligibility. Most speaker anonymization systems disentangle the speaker representation from the original speech and achieve anonymization by averaging or modifying the speaker representation. However, the anonymized speech is subject to reduction in pseudo speaker distinctiveness, speech quality and intelligibility for out-of-distribution speaker. To solve this issue, we propose SALT, a Speaker Anonymization system based on Latent space Transformation. Specifically, we extract latent features by a self-supervised feature extractor and randomly sample multiple speakers and their weights, and then interpolate the latent vectors to achieve speaker anonymization.  Meanwhile, we explore the extrapolation method to further extend the diversity of pseudo speakers. Experiments on Voice Privacy Challenge dataset show our system achieves a state-of-the-art distinctiveness metric while preserving speech quality and intelligibility. Our code and demo is availible at github\footnote{\url{https://github.com/BakerBunker/SALT}}.
\blfootnote{This work was supported in part by Shanghai Pudong New Area Science and Technology Development Fund under Grant No. PKS2022-04.}

% Speaker anonymization aims to conceal a speaker's identity without degrading speech quality and intelligibility. Current mainstream speaker anonymization systems disentangle the speaker-related representation from the original speech signal and achieve anonymization by averaging or modifying speaker-related representation.
% However, the resulting anonymized speech is subject to reduction in pseudo speaker diversity and degradation in speech quality and intelligibility for unseen speaker distribution.
% To solve this issue, we propose \modelname{}, a speaker anonymization system based on vector-matching and interpolation. 
% We extract latent vectors by the self-supervised feature extractor and randomly sample multiple speakers and their weights, then interpolate the latent vectors to achieve speaker anonymization. Meanwhile, we explore the extrapolation method to extend the distinctiveness of pseudonymous speakers to outperform the existing baseline.
% Experimental results demonstrate our system can protect privacy effectively and achieve a positive Gain of Voice Distinctiveness ($G_{\mathrm{VD}}$) index while maintaining speech quality and intelligibility. 

\end{abstract}
\begin{keywords}
voice privacy, speaker anonymization, voice conversion, speech synthesis
\end{keywords}

\begin{figure*}[h]
\centering
\vspace{-10pt}
\includegraphics[width=0.8\linewidth]{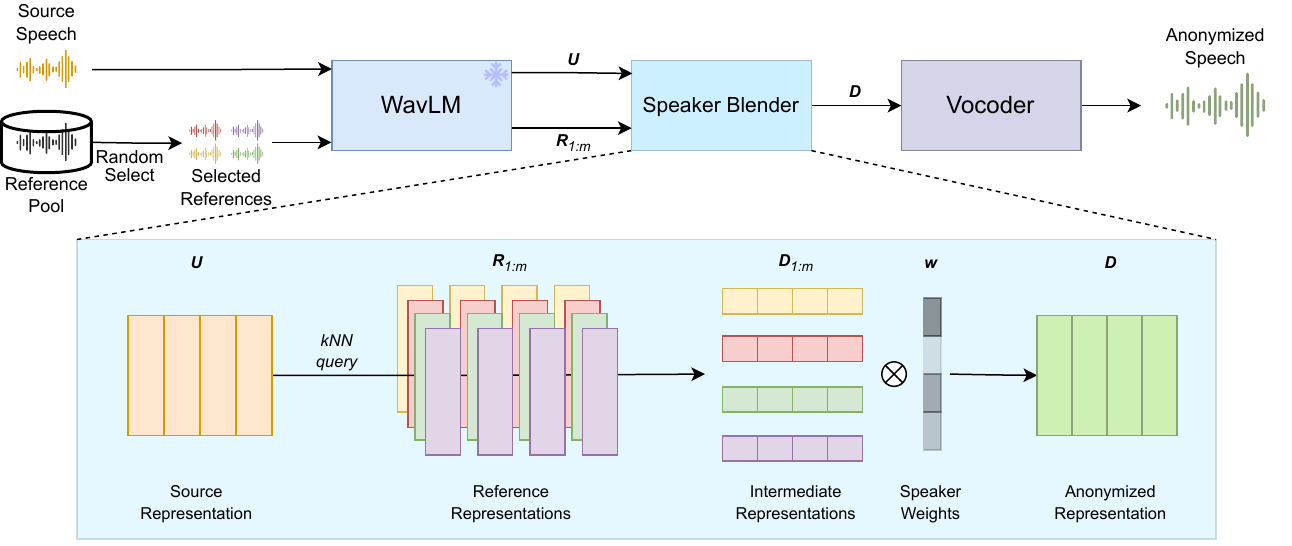}
\caption{Illustration of \modelname{} Pipeline. The WavLM encoder is frozen during training and the blue box is the detail of our proposed speaker blender. Vocoder is used to reconstruct the anonymized speech.}
\label{fig:pipeline}
\end{figure*}
\section{Introduction}
\vspace{-5pt}
\label{sec:intro}

Speech data on the Internet are exponentially proliferating because of the widely usage of social media. However, current voice biometrics has the capability to extract various personal sensitive information from a given speech signal, including the speaker's identity, age, gender, and even health state~\cite{8316845}.
If a malicious attacker has access to an individual's voice data, there is a risk of compromise of sensitive user information.
In order to mitigate this concern, the European Union has recently introduced new regulations such as the General Data Protection Regulation (GDPR), which aims to strengthen privacy preservation and protect personal speech data~\cite{EuropeanParliament2016a}.
Therefore, a new challenge arises regarding how to eliminate identity information from speech while preserving the content of the speech data.

\textit{Speaker anonymization}, a user-centric voice privacy solution, aims at concealing a speaker's identity without degrading intelligibility and naturalness.
Currently, speaker anonymization is still in its infancy. To accelerate the advancements in this field, the VoicePrivacy Challenge (VPC) was held in 2020 and 2022, focusing on developing privacy preservation solutions for speech technology~\cite{Tomashenko2022TheV2}. During these two challanges, numerous systems for speaker anonymization were proposed, leading to significant progress in speaker anonymization techniques.

% The evaluation of the VPC reveals that the majority of speaker anonymization systems adhere to one of the established baselines. Consequently, these approaches
Most of the approaches in VPC can be categorized into two distinct groups: (1) signal-processing based voice transformation and (2) x-vector based voice conversion. Signal-processing based speaker anonymization system does not need training data and directly modifies speech characteristics such as the pitch, spectral envelope, and time scaling~\cite{Mawalim2022SpeakerAB}. However, attackers might be able to restore the original speech after a reasonable number of attempts, due to the scope of physical shifts for speech signals is limited.
State-of-the-art anonymization systems are inspired by disentangling the speaker-related representation from the original speech signal by neural network, this method is also widely used in neural voice conversion.
These systems use an x-vector extracted from a pre-trained automatic speaker verification (ASV) model as speaker-related representation, and then eliminate speaker-specific characteristics by averaging or modifying candidate x-vectors~\cite{Meyer2022SpeakerAW,Fang2019SpeakerAU,Gupta2020DesignOV}. Subsequent works have explored various approaches to improve the anonymization performance. Mawalim \et utilized singular value decomposition (SVD) to improve the anonymization performance~\cite{DBLP:conf/interspeech/MawalimGKU20}. Champion \et, suggested F0 as the critical factor and modified it to achieve anonymization~\cite{DBLP:journals/corr/abs-2101-08478}, and Turner \et employed Gaussian Mixed Model (GMM) to sample pseudo speakers in a principal components analysis (PCA) compact space where the original distribution of cosine distances between x-vectors is retained~\cite{DBLP:journals/corr/abs-2010-13457}.

Despite the effectiveness of these approaches, there remain many aspects for improvement including increasing the distinctiveness of anonymization speech and dealing with more powerful attack scenarios.
Firstly, due to the restricted sampling space around the average speaker, the current approches suffer from a lack of timbre diversity among pseudo speakers. Moreover, since these models are trained on seen speakers, they encounter significant degradation in speech quality  when the candidate speaker vectors are out-of-distribution. In other words, when randomly generated pseudo speaker vectors are far from the distribution of the training set, the resulting speech can exhibit a substantial decline in quality. Finally, x-vector based methods involves disentangled content and speaker information using automatic speech recognition (ASR) and speaker verification (SV) models, which inevitably leads to information loss or leakage, resulting in a degradation of speaker distinctiveness and audio quality.

To address these problems, this study proposes a vector-matching and latent transformation based speaker anonymization system, which is inspired by kNN-VC, a voice conversion (VC) model based on the K-nearest neighbors (KNN) algorithm~\cite{baas2023knnvc}. Unlike most existing voice conversion systems, kNN-VC uses WavLM~\cite{DBLP:journals/jstsp/ChenWCWLCLKYXWZ22}, a self-supervised model, to extract latent representation, which is relatively more robust than supervised models according to previous research ~\cite{DBLP:conf/nips/HendrycksMKS19}, thereby reducing the degradation in speech quality and intelligibility caused by out-of-distribution (OOD) scenarios. Leveraging the speech latent space representations obtained from the pre-trained model, we introduce an latent transformation based speaker blender. It randomly samples multiple speakers and their weights, then interpolate the latent representation to achieve speaker anonymization. Furthermore, we explore the extrapolation method for speaker anonymization, extending the diversity of pseudo speakers beyond the existing baseline. Finally, the anonymized latent features are transformed into speech by a modified HiFiGAN vocoder~\cite{DBLP:conf/nips/KongKB20}. Experiments demonstrate that our proposed anonymization method outperforms the NWPU-ASLP system~\cite{DBLP:journals/corr/abs-2209-11969}, which ranked the first in VPC 2022~\cite{Tomashenko2022TheV2}, in all anonymization related metrics, particularly in terms of speaker distinctiveness.

% \textcolor{blue}{
% This study was inspired by kNN-VC, a voice conversion (VC) model based on K-nearest neighbors (KNN) for matching latent space vectors, it proposes a vector-matching-based speaker anonymization model. Unlike existing anonymization systems, our approach utilizes a pre-trained model, allowing us to obtain relatively robust content and speaker representations, thereby reducing the degradation in speech quality and intelligibility caused by out-of-distribution (OOD) scenarios. Leveraging the speech latent space representations obtained from the pre-trained model, we introduce an interpolation-based system. It involves randomly sampling from multiple different speakers and assigning random weights to achieve speaker anonymization. Furthermore, we explore the extrapolation method for speaker anonymization, extending the diversity of pseudonymous speakers beyond the existing baseline. Finally, the interpolated latent space features are transformed into speech using the modified HiFiGAN vocoder. Experiment results demonstrate that our proposed anonymization method outperforms the baseline in all evaluation metrics, particularly in terms of speaker diversity.}

\section{Method}

The architecture of our proposed model, as depicted in the upper of Fig.\ref{fig:pipeline}, can be divided into three distinct components: a pre-trained self-supervised latent encoder, a speaker blender, and a vocoder. 
The self-supervised latent encoder encodes source and reference waveforms into latent representations. The speaker blender is used to randomly interpolate the features of reference speeches and get the anonymized latent representation of source speech. Ultimately, the vocoder converts the anonymized latent representation into an anonymized speech.

\subsection{Encoder}

%we use ...; compare to existing anonymization ...
To extract the joint representation of linguistic content and speaker information, we utilize a pre-trained WavLM model as our feature extractor.
% In the proposed method of this paper, we utilize pre-trained WavLM to extract a joint representation of content and speaker information. 
Previous studies have demonstrated that the distance between latent representations of the same phoneme is greater than different phonemes~\cite{DBLP:journals/jstsp/DunbarHD22}. Additionally, for different speakers, the pre-trained representations also tend to cluster based on speaker characteristics~\cite{DBLP:journals/jstsp/ChenWCWLCLKYXWZ22}. Therefore,  leveraging these properties, we can interpolate the latent representation to achieve speaker anonymization. 
The goal of Encoder is to extract features of source speech $\mathbf{U}$ and features of reference speeches $\mathbf{R}_{spk}$, denoted as:
\begin{alignat}{3}
    \mathbf{U}&=[\mathbf{u}_i]_{i=1}^T&&=\mathrm{WavLM}(\mathbf{x}) \\
    \mathbf{R}_{spk}&=[\mathbf{r}_i]_{i=1}^{T_{spk}}&&=\mathrm{WavLM}(\mathbf{x}_{spk}) \ where\ spk\in[0, N)
\end{alignat}
where $\mathbf{x}$ and $\mathbf{x}_{spk}$ represent the source speech and the reference speech of speaker $spk$, respectively and $spk\in[0, N)$ and $N$ represent the number of reference speakers.  $T$, $T_{spk}$ represents the source frame length and reference frame length of speaker $spk$.

\subsection{Speaker Blender}
\label{sec:matcher}
\subsubsection{Pseudo Speaker Generator}

In order to transform the latent space representation of the original speech into an anonymized latent space representation, we design a speaker blender, the structure illustrated as the lower of Fig.\ref{fig:pipeline}.
During the generation process of pseudo speakers, a set $\mathbf{M}\subseteq [0,N)$ with $m$ speakers is randomly selected from the reference speaker pool. For each speaker $spk$, the k-nearest neighbors (kNN) algorithm is used to obtain the closest representations $\mathbf{D}_{spk}$ from the target reference speaker feature set $\mathbf{R}_{spk}$. This step can be described as:
\begin{equation}
    \mathbf{D}_{spk}=[\operatorname{kNN}(\mathbf{u}_i,\mathbf{R}_{spk},k)]_{i=1}^T \qquad where\ spk\in \mathbf{M}
\end{equation}
where $\mathrm{kNN}(\mathbf{x},\mathbf{Y},k)$ means find $k$ nearest vectors to vector $\mathbf{x}$ in set $\mathbf{Y}$.

To generate the pseudo speaker, we need to perform random interpolation on the representations of these target speakers. Let randomly generated speaker weight vector $\mathbf{w}=(w_1,w_2,\dots,w_{}) $, and the weight vector is calculated from:
\begin{equation}
    \mathbf{w}_{spk}=\begin{cases}
        \mathcal{N}(0,1), &\text{if} \;spk \in \mathbf{M}.\\
        -\infty, &\text{otherwise.}
    \end{cases}
\end{equation}
Then we apply the softmax function to $w$ to constrain the sum of weights $\sum_{i=1}^m w_i$ to 1. 

Finally, the target pseudo speaker representation $D$ is obtained by performing a weighted sum of the representations from each speaker, denoted as:
\begin{equation}
    D=\sum w_{spk}D_{spk} \qquad  where \ spk\in M
\end{equation}
where $w_k$ denotes the weight of speaker $k$.

In scenarios where voice quality is more important than anonymity, we can preserve the feature of source speaker $S$ in the final latent feature $D$ with a preservation factor $p$, denoted as:
\begin{equation}
    D=pS+(1-p)\sum w_{spk}D_{spk} \quad  where \ spk\in M
\end{equation}

% Through these steps, we get $w$, a unique descriptor vector for a pseudo speaker, so we can easily memorize this vect

\subsubsection{Speaker Extrapolation}
In recent speaker anonymization methods, many of them rely on random perturbations around the average speaker vector, leading to insufficient or approximate pseudo speakers and resulting in the problem of pseudo speakers being too similar to each other. In our proposed anonymization system, we introduce speaker extrapolation to alleviate this issue. Specifically, after obtaining the weight vector $\mathbf{w}$ of reference speakers, we scale the vector to expand its value range from $[0,1]$ to $[-s/m,(s+1)-s/m]$, where $s$ denotes the scale factor, the scaling function is denoted as:
\begin{equation}
    \mathbf{w}^\prime=\mathbf{w}*(s+1)-s/m
\end{equation}
where $\mathbf{w}^\prime$ denotes extrapolated speaker weight vector.
By scaling the weight vector of the pseudo speakers, we can obtain more widely distributed pseudo speaker representations, which leads to more diverse anonymous speakers.

\subsection{Vocoder}

To reconstruct the latent representations into anonymized speech signals, % we adopt a similar approach to kNN-VC and employ the architecture of HiFiGAN-V1. 
Inspired by kNN-VC, we leverage the powerful HiFiGAN-V1 as our vocoder~\cite{DBLP:conf/nips/KongKB20}.
Prior to training, a crucial step involved conducting pre-matching on the training set. This process entails selecting a representative portion of audio from each speaker and extracting latent space representations, which served as the reference set. Subsequently, for each audio sample within the training set, we perform kNN matching to reconstruct their respective latent space features. During training, we use the pre-matched feature-audio pairs for training.

\begin{table*}[]
  \centering
  \footnotesize
  \renewcommand{\tabcolsep}{0.11cm}
    \caption{EER achieved by $ASV_{eval}^{anon}$ on data processed by our anonymization method vs. EER achieved by baseline B1.a or NWPU-ASLP and original (Orig). Our proposed model is denoted as [B or L]-S$x$-P$x$, where B or L means WavLM-Base or WavLM-Large encoder, S$x$ means the scale factor $s=x$, P$x$ means the preservation factor $p=x$.}
\begin{tabular}{|ccc|c|cc|c|cccc|}
\hline
\multicolumn{1}{|c|}{\multirow{2}{*}{\textbf{Dataset}}} & \multicolumn{1}{l|}{\multirow{2}{*}{\textbf{Gender}}} & \multicolumn{1}{l|}{\multirow{2}{*}{\textbf{Weight}}} & \multicolumn{1}{l|}{\multirow{2}{*}{\textbf{Orig}}} & \multicolumn{2}{c|}{\textbf{VPC Baseline}} & \textbf{VPC Top1} & \multicolumn{4}{c|}{\textbf{SALT (Proposed)}} \\ \cline{5-11} 
\multicolumn{1}{|c|}{} & \multicolumn{1}{l|}{} & \multicolumn{1}{l|}{} & \multicolumn{1}{l|}{} & \multicolumn{1}{c|}{\textbf{B1.a}} & \textbf{B1.b} & \textbf{NWPU-ASLP} & \multicolumn{1}{c|}{\textbf{B-S0-P0}} & \multicolumn{1}{c|}{\textbf{B-S1-P0}} & \multicolumn{1}{c|}{\textbf{B-S0-P0.2}} & \textbf{L-S0-P0} \\ \hline
\multicolumn{1}{|c|}{\multirow{2}{*}{LibriSpeech-dev}} & \multicolumn{1}{c|}{female} & 0.25 & 8.67 & \multicolumn{1}{c|}{17.76} & 19.03 & 26.28 & \multicolumn{1}{c|}{44.32} & \multicolumn{1}{c|}{54.4} & \multicolumn{1}{c|}{\textbf{55.97}} & 52.70 \\
\multicolumn{1}{|c|}{} & \multicolumn{1}{c|}{male} & 0.25 & 1.24 & \multicolumn{1}{c|}{6.37} & 5.59 & 23.45 & \multicolumn{1}{c|}{49.38} & \multicolumn{1}{c|}{45.5} & \multicolumn{1}{c|}{49.53} & \textbf{50.31} \\ \hline
\multicolumn{1}{|c|}{\multirow{2}{*}{VCTK-dev (diff)}} & \multicolumn{1}{c|}{female} & 0.20 & 2.86 & \multicolumn{1}{c|}{12.46} & 8.25 & 40.31 & \multicolumn{1}{c|}{33.24} & \multicolumn{1}{c|}{39.19} & \multicolumn{1}{c|}{\textbf{44.30}} & 43.46 \\
\multicolumn{1}{|c|}{} & \multicolumn{1}{c|}{male} & 0.20 & 1.44 & \multicolumn{1}{c|}{9.33} & 6.01 & 27.77 & \multicolumn{1}{c|}{42.38} & \multicolumn{1}{c|}{46.35} & \multicolumn{1}{c|}{\textbf{52.06}} & 32.26 \\ \hline
\multicolumn{1}{|c|}{\multirow{2}{*}{VCTK-dev (comm)}} & \multicolumn{1}{c|}{female} & 0.05 & 2.62 & \multicolumn{1}{c|}{13.95} & 9.01 & 35.76 & \multicolumn{1}{c|}{38.95} & \multicolumn{1}{c|}{\textbf{47.67}} & \multicolumn{1}{c|}{44.19} & 46.51 \\
\multicolumn{1}{|c|}{} & \multicolumn{1}{c|}{male} & 0.05 & 1.43 & \multicolumn{1}{c|}{13.11} & 9.40 & 37.89 & \multicolumn{1}{c|}{45.58} & \multicolumn{1}{c|}{46.44} & \multicolumn{1}{c|}{\textbf{53.56}} & 36.47 \\ \hline
\multicolumn{3}{|c|}{Weighted average dev} & 3.54 & \multicolumn{1}{c|}{11.74} & 9.93 & 31.73 & \multicolumn{1}{c|}{42.78} & \multicolumn{1}{c|}{46.79} & \multicolumn{1}{c|}{\textbf{50.53}} & 45.05 \\ \hline
\multicolumn{1}{|c|}{\multirow{2}{*}{LibriSpeech-test}} & \multicolumn{1}{c|}{female} & 0.25 & 7.66 & \multicolumn{1}{c|}{12.04} & 9.49 & 22.08 & \multicolumn{1}{c|}{44.71} & \multicolumn{1}{c|}{\textbf{44.89}} & \multicolumn{1}{c|}{37.77} & 39.96 \\
\multicolumn{1}{|c|}{} & \multicolumn{1}{c|}{male} & 0.25 & 1.11 & \multicolumn{1}{c|}{8.91} & 7.80 & 19.15 & \multicolumn{1}{c|}{\textbf{53.45}} & \multicolumn{1}{c|}{51.67} & \multicolumn{1}{c|}{38.31} & 48.11 \\ \hline
\multicolumn{1}{|c|}{\multirow{2}{*}{VCTK-test (diff)}} & \multicolumn{1}{c|}{female} & 0.20 & 4.89 & \multicolumn{1}{c|}{16.00} & 10.91 & 40.64 & \multicolumn{1}{c|}{\textbf{56.94}} & \multicolumn{1}{c|}{56.07} & \multicolumn{1}{c|}{42.54} & 32.10 \\
\multicolumn{1}{|c|}{} & \multicolumn{1}{c|}{male} & 0.20 & 2.07 & \multicolumn{1}{c|}{10.05} & 7.52 & 38.81 & \multicolumn{1}{c|}{34.16} & \multicolumn{1}{c|}{24.28} & \multicolumn{1}{c|}{44.51} & \textbf{51.55} \\ \hline
\multicolumn{1}{|c|}{\multirow{2}{*}{VCTK-test (comm)}} & \multicolumn{1}{c|}{female} & 0.05 & 2.89 & \multicolumn{1}{c|}{17.34} & 15.32 & 40.46 & \multicolumn{1}{c|}{48.27} & \multicolumn{1}{c|}{\textbf{52.89}} & \multicolumn{1}{c|}{39.90} & 37.28 \\
\multicolumn{1}{|c|}{} & \multicolumn{1}{c|}{male} & 0.05 & 1.13 & \multicolumn{1}{c|}{9.89} & 8.19 & 38.70 & \multicolumn{1}{c|}{42.09} & \multicolumn{1}{c|}{\textbf{43.50}} & \multicolumn{1}{c|}{37.57} & 36.72 \\ \hline
\multicolumn{3}{|c|}{Weighted average test} & 3.79 & \multicolumn{1}{c|}{11.81} & 9.18 & 30.15 & \multicolumn{1}{c|}{\textbf{47.28}} & \multicolumn{1}{c|}{45.03} & \multicolumn{1}{c|}{39.61} & 42.45 \\ \hline
\end{tabular}

  \label{tab:eer}
  \vspace{-10pt}
\end{table*}

\section{Experiments}

To evaluate the effectiveness of our latent transformation based speaker anonymization system, we follow the VPC 2022 evaluation scheme to conduct experiments~\cite{Tomashenko2022TheV2}. Our speech demos are placed at demo-site.\footnote{\url{https://bakerbunker.github.io/SALT}}.

\subsection{Datasets}
To ensure consistency, training datasets follow the same configuration of the VPC 2022. Thus, the LibriSpeech corpus is used for training the WavLM-base and we use the LibriSpeech-train-clean-100 dataset as the vocoder training set~\cite{DBLP:conf/icassp/PanayotovCPK15}. Furthermore, to facilitate a comprehensive comparison among different latent extractors, we extend the scope from WavLM-base to WavLM-large. In this expansion, a significantly broader dataset consisting of 94,000 hours is employed to train the WavLM-large model, allowing a deeper exploration of the capabilities and performance of the latent extractor.

As for evaluation, we use the official VPC development and test sets. These two sets contain several female and male speakers from the LibriSpeech~\cite{DBLP:conf/icassp/PanayotovCPK15} and VCTK~\cite{yamagishi2019vctk} corpus. Following the VPC guidance, we fine-tune the official ASV model, denoted as $ASV_{eval}^{anon}$, by leveraging our anonymized speech data in different hyper-parameters to create a simulated Semi-informed attack condition.
%如果用不finetune的asr再加一句说明
% In the training process of baseline, we use the same dataset as WavLM-Base.
% In our study, we adopt the widely used LibriSpeech~\cite{DBLP:conf/icassp/PanayotovCPK15} and VCTK~\cite{yamagishi2019vctk} datasets as our training and testing datasets which is similar to VoicePrivacy Challenge 2022.
% The LibriSpeech dataset consists of 960 hours of speech data, while the VCTK dataset contains 44 hours of speech data. According to Chen \et~\cite{DBLP:journals/jstsp/ChenWCWLCLKYXWZ22}, WavLM-base trained on the LibriSpeech dataset, while WavLM-Large employed 94,000 hours of data for training. For the part of the vocoder, we utilized the LibriSpeech-train-clean-100 dataset as our training set. In the training process of baseline, we use the same dataset as WavLM-Base. In the evaluation, we employed the test set and dev set from both LibriSpeech and VCTK datasets. For VCTK dataset, we split it to two subsets: common and different. The common subset comprises utterances \#1-24 in the VCTK corpus that are identical for all speakers. This choice is intended to support subjective evaluation of speaker verifiability in a text-dependent manner.

\subsection{Model Setup}

In the encoder part, we employ pre-trained WavLM-Large and WavLM-Base models\footnote{\url{https://github.com/microsoft/unilm/tree/master/wavlm}} to extract the latent space representations of source and reference speeches. For WavLM-Large, we follow the setup of kNN-VC and use the 6th layer feature as the latent space representation, while for WavLM-Base, we use the feature of the 3rd layer, which is the same proportion of layers as WavLM-Large and achieve the best reconstruction performance in our experiments.

In the speaker blender part, for the kNN matching, we set the value of $k$ to 4 and utilize cosine similarity as the distance metric. For the interpolation method, we randomly select 50 speakers from the LibriSpeech train-clean-100 dataset. For each selected speaker, we  randomly choose 50 audio samples to extract features. The number of random reference speakers $m$ is set to 4. The preservation factor $p$ is set to $0$ and $0.2$. For the extrapolation method, we experiment with two different scale factors: $0$ and $1$, where a scale factor of $0$ indicates no extrapolation involved.

We train the vocoder for the WavLM-Base feature using the same hyperparameters as the kNN-VC. As for the vocoder of WavLM-Large features,  we utilize the model parameters from the open-source kNN-VC repository\footnote{\url{https://github.com/bshall/knn-vc}}. 
% For the vocoder which transforms WavLM-Large features to the waveform, we utilized the model parameters from the open-source kNN-VC repository. For the vocoder for WavLM-Base features, we employed the same data, optimizer, training steps, batch size, and other hyperparameters as the former.

% In subsequent experiments, the model is denoted as [B or L]-S$x$-P$x$, where B or L means WavLM-Base or WavLM-Large encoder, S$x$ means the scale factor $s=x$, P$x$ means the preservation factor $p=x$.

\subsection{Baselines}

We compare the proposed system to the primary baselines of the VPC 2022 B1.a and B1.b~\cite{Tomashenko2022TheV2} and NWPU-ASLP anonymization system which ranked first in VPC 2022~\cite{DBLP:journals/corr/abs-2209-11969}.

\textit{VPC baseline systems}: The anonymization process of the baseline system involves three steps. Firstly, the system utilizes a speaker verification model to extract the x-vector as a speaker information feature, an automatic speech recognition (ASR) model to extract bottleneck features (BN) as linguistic content features, also extract fundamental frequency (F0) as intonation information. 
In the second step, a set of candidate x-vectors are selected from an x-vectors pool relying on the farthest cosine distance metric of the original speaker, these x-vectors are then averaged to obtain the anonymized x-vector. Finally, the acoustic model and vocoder are employed to synthesize the anonymized speech based on the anonymized x-vector, BN, and F0.

\textit{NWPU-ASLP system}~\cite{DBLP:journals/corr/abs-2209-11969}: The NWPU-ASLP system preserves a pseudo speaker id in the look-up-table to generate pseudo speaker embedding and utilizes averaged embedding as the condition for pseudo speaker embedding to produce the final anonymized embedding. Subsequently, the acoustic model generates anonymized mel-spectrograms based on the obtained anonymized embedding. 
Finally, a vocoder retrained with ground truth alignment data is employed to reconstruct the anonymized mel-spectrogram into speech signals.

\subsection{Evaluation Metrics}

In order to measure the proposed anonymization system in privacy protection and utility, five metrics are employed:
\begin{itemize}

\item \textbf{Equal Error Rate (EER)} relies on an ASV model provided by the VPC 2022 and assesses the privacy protection ability of the anonymization system. A higher EER indicates better performance in anonymizing the speaker, as it implies increased difficulty in correctly identifying the speaker.

\item \textbf{Gain of voice distinctiveness ($\mathrm{G_{VD}}$)} quantifies the degree of speaker distinctiveness before and after anonymization~\cite{DBLP:conf/interspeech/NoeBMTNE20,DBLP:journals/csl/NoeNEPBTM22}. It is computed by analyzing the diagonal dominance proportion of the embedding matrix for the original and anonymized speakers. A higher Gain of voice distinctiveness ($\mathrm{G_{VD}}$) indicates better preservation of speaker distinctiveness.

\item \textbf{Pitch Correlation ($\rho^\mathrm{F0}$)} evaluates the preservation of speech intonation before and after anonymization. The pitch correlation metric $\rho^\mathrm{F0}$ is the Pearson correlation between the pitch sequences, estimated according to ~\cite{Hirst2007APP}, of original and anonymized utterances. A higher correlation ($\rho^\mathrm{F0}$) indicates better preservation of speech intonation.

\item \textbf{Word Error Rate (WER)} relies on a pre-trained ASR model and measures the objective intelligibility of the anonymized speech. We use the Whisper Large speech recognition system\footnote{\url{https://github.com/openai/whisper}}, as this model leverages a large speech corpus, ensuring its stability under different audio conditions and approaching human-level comprehension~\cite{DBLP:journals/corr/abs-2212-04356}. We also use U2++~\cite{DBLP:journals/corr/abs-2106-05642} adopted from WeNet toolkit~\cite{DBLP:conf/interspeech/YaoWWZYYPCXL21}\footnote{\url{https://github.com/wenet-e2e/wenet/tree/main/examples/gigaspeech/s0}} which is pretrained on GigaSpeech dataset~\cite{DBLP:conf/interspeech/ChenCWDZWSPTZJK21}, and TDNN-F~\cite{DBLP:conf/interspeech/PoveyCWLXYK18} with the same train and finetune procedure with VPC 2022. A lower WER indicates better intelligibility of the anonymized speech.
%加vpc和u2++还需要额外说明

\item \textbf{Mean Opinion Score (MOS)} is conducted to gather subjective evaluations of the anonymization system. The MOS assessment rate the speech from the naturalness aspect, where a panel of listeners rates the audio on a scale of 1 to 5. Higher scores indicate better subjective perception. The experiment collected ratings from a group of 10 participants to obtain a comprehensive evaluation of the anonymization system's subjective performance.

\end{itemize}

\begin{table}[ht]
  
  \renewcommand\arraystretch{1.2}
  \renewcommand{\tabcolsep}{0.14cm}
  \centering
  \footnotesize
  \vspace{-10pt}
    \caption{Pitch correlation $\rho_\mathrm{F0}$ achieved on data processed by B1.a, NWPU-ASLP and our anonymized results.}
  \resizebox{0.46\textwidth}{!}{
\begin{tabular}{|cc|cc|c|cccc|}
\hline
\multicolumn{1}{|c|}{\multirow{2}{*}{\textbf{Dataset}}} & \multicolumn{1}{l|}{\multirow{2}{*}{\textbf{Gender}}} & \multicolumn{2}{c|}{\textbf{VPC Baseline}} & \textbf{VPC Top1} & \multicolumn{4}{c|}{\textbf{SALT (Proposed)}} \\ \cline{3-9} 
\multicolumn{1}{|c|}{} & \multicolumn{1}{l|}{} & \multicolumn{1}{c|}{\textbf{B1.a}} & \textbf{B1.b} & \textbf{NWPU-ASLP} & \multicolumn{1}{c|}{\textbf{B-S0-P0}} & \multicolumn{1}{c|}{\textbf{B-S1-P0}} & \multicolumn{1}{c|}{\textbf{B-S0-P0.2}} & \textbf{L-S0-P0} \\ \hline
\multicolumn{1}{|c|}{\multirow{2}{*}{LibriSpeech-dev}} & female & \multicolumn{1}{c|}{0.77} & \textbf{0.84} & 0.71 & \multicolumn{1}{c|}{0.80} & \multicolumn{1}{c|}{0.82} & \multicolumn{1}{c|}{\textbf{0.84}} & 0.83 \\
\multicolumn{1}{|c|}{} & male & \multicolumn{1}{c|}{0.73} & 0.76 & 0.69 & \multicolumn{1}{c|}{0.78} & \multicolumn{1}{c|}{0.74} & \multicolumn{1}{c|}{\textbf{0.79}} & 0.78 \\ \hline
\multicolumn{1}{|c|}{\multirow{2}{*}{VCTK-dev (dif)}} & female & \multicolumn{1}{c|}{0.84} & \textbf{0.87} & 0.76 & \multicolumn{1}{c|}{0.84} & \multicolumn{1}{c|}{0.85} & \multicolumn{1}{c|}{0.86} & 0.85 \\
\multicolumn{1}{|c|}{} & male & \multicolumn{1}{c|}{0.78} & 0.76 & 0.71 & \multicolumn{1}{c|}{0.78} & \multicolumn{1}{c|}{0.78} & \multicolumn{1}{c|}{\textbf{0.80}} & 0.79 \\ \hline
\multicolumn{1}{|c|}{\multirow{2}{*}{VCTK-dev (com)}} & female & \multicolumn{1}{c|}{0.79} & \textbf{0.84} & 0.71 & \multicolumn{1}{c|}{0.80} & \multicolumn{1}{c|}{0.82} & \multicolumn{1}{c|}{0.83} & 0.82 \\
\multicolumn{1}{|c|}{} & male & \multicolumn{1}{c|}{0.72} & 0.72 & 0.67 & \multicolumn{1}{c|}{0.76} & \multicolumn{1}{c|}{0.74} & \multicolumn{1}{c|}{\textbf{0.78}} & 0.76 \\ \hline
\multicolumn{2}{|c|}{Weighted average dev} & \multicolumn{1}{c|}{0.77} & 0.80 & 0.72 & \multicolumn{1}{c|}{0.80} & \multicolumn{1}{c|}{0.79} & \multicolumn{1}{c|}{\textbf{0.82}} & 0.80 \\ \hline
\multicolumn{1}{|c|}{\multirow{2}{*}{LibriSpeech-test}} & female & \multicolumn{1}{c|}{0.77} & \textbf{0.85} & 0.72 & \multicolumn{1}{c|}{0.83} & \multicolumn{1}{c|}{0.78} & \multicolumn{1}{c|}{0.84} & 0.84 \\
\multicolumn{1}{|c|}{} & male & \multicolumn{1}{c|}{0.69} & 0.72 & 0.64 & \multicolumn{1}{c|}{0.74} & \multicolumn{1}{c|}{0.73} & \multicolumn{1}{c|}{0.74} & \textbf{0.75} \\ \hline
\multicolumn{1}{|c|}{\multirow{2}{*}{VCTK-test (dif)}} & female & \multicolumn{1}{c|}{\textbf{0.87}} & \textbf{0.87} & 0.77 & \multicolumn{1}{c|}{0.83} & \multicolumn{1}{c|}{0.81} & \multicolumn{1}{c|}{\textbf{0.87}} & 0.84 \\
\multicolumn{1}{|c|}{} & male & \multicolumn{1}{c|}{0.79} & 0.77 & 0.71 & \multicolumn{1}{c|}{0.79} & \multicolumn{1}{c|}{0.77} & \multicolumn{1}{c|}{\textbf{0.80}} & \textbf{0.80} \\ \hline
\multicolumn{1}{|c|}{\multirow{2}{*}{VCTK-test (com)}} & female & \multicolumn{1}{c|}{0.79} & \textbf{0.85} & 0.72 & \multicolumn{1}{c|}{0.80} & \multicolumn{1}{c|}{0.77} & \multicolumn{1}{c|}{0.84} & 0.82 \\
\multicolumn{1}{|c|}{} & male & \multicolumn{1}{c|}{0.70} & 0.71 & 0.65 & \multicolumn{1}{c|}{0.74} & \multicolumn{1}{c|}{0.72} & \multicolumn{1}{c|}{0.74} & \textbf{0.75} \\ \hline
\multicolumn{2}{|c|}{Weighted average test} & \multicolumn{1}{c|}{0.77} & \textbf{0.80} & 0.70 & \multicolumn{1}{c|}{0.79} & \multicolumn{1}{c|}{0.77} & \multicolumn{1}{c|}{\textbf{0.80}} & \textbf{0.80} \\ \hline
\end{tabular}
}

  \label{tab:f0}
  \vspace{-10pt}
\end{table}

\begin{table}[htb]
  \renewcommand\arraystretch{1.4}
  \renewcommand{\tabcolsep}{0.14cm}
  \centering
  \footnotesize
    \caption{Gain of voice distinctiveness $G_\mathrm{VD}$ achieved on data processed by B1.a, NWPU-ASLP and our anonymized results.}
  \resizebox{0.46\textwidth}{!}{
\begin{tabular}{|cc|cc|c|cccc|}
\hline
\multicolumn{1}{|c|}{\multirow{2}{*}{\textbf{Dataset}}} & \multirow{2}{*}{\textbf{Gender}} & \multicolumn{2}{c|}{\textbf{VPC Baseline}} & \textbf{VPC Top1} & \multicolumn{4}{c|}{\textbf{SALT (Proposed)}} \\ \cline{3-9} 
\multicolumn{1}{|c|}{} &  & \multicolumn{1}{c|}{\textbf{B1.a}} & \textbf{B1.b} & \textbf{NWPU-ASLP} & \multicolumn{1}{c|}{\textbf{B-S0-P0}} & \multicolumn{1}{c|}{\textbf{B-S1-P0}} & \multicolumn{1}{c|}{\textbf{B-S0-P0.2}} & \textbf{L-S0-P0} \\ \hline
\multicolumn{1}{|c|}{\multirow{2}{*}{LibriSpeech-dev}} & female & \multicolumn{1}{c|}{-9.15} & -4.92 & -21.35 & \multicolumn{1}{c|}{0.11} & \multicolumn{1}{c|}{\textbf{0.15}} & \multicolumn{1}{c|}{-0.03} & 0.06 \\
\multicolumn{1}{|c|}{} & male & \multicolumn{1}{c|}{-8.94} & -6.38 & -18.66 & \multicolumn{1}{c|}{\textbf{-0.08}} & \multicolumn{1}{c|}{-0.11} & \multicolumn{1}{c|}{\textbf{-0.08}} & -0.10 \\ \hline
\multicolumn{1}{|c|}{\multirow{2}{*}{VCTK-dev (dif)}} & female & \multicolumn{1}{c|}{-8.82} & -5.94 & -13.96 & \multicolumn{1}{c|}{\textbf{0.53}} & \multicolumn{1}{c|}{0.42} & \multicolumn{1}{c|}{0.29} & 0.46 \\
\multicolumn{1}{|c|}{} & male & \multicolumn{1}{c|}{-12.61} & -9.38 & -20.72 & \multicolumn{1}{c|}{0.09} & \multicolumn{1}{c|}{\textbf{0.29}} & \multicolumn{1}{c|}{0.12} & 0.23 \\ \hline
\multicolumn{1}{|c|}{\multirow{2}{*}{VCTK-dev (com)}} & female & \multicolumn{1}{c|}{-7.56} & -4.17 & -17.18 & \multicolumn{1}{c|}{0.19} & \multicolumn{1}{c|}{0.16} & \multicolumn{1}{c|}{0.09} & \textbf{0.23} \\
\multicolumn{1}{|c|}{} & male & \multicolumn{1}{c|}{-10.37} & -6.99 & -19.71 & \multicolumn{1}{c|}{-0.07} & \multicolumn{1}{c|}{\textbf{0.08}} & \multicolumn{1}{c|}{0.02} & 0.06 \\ \hline
\multicolumn{2}{|c|}{Weighted average dev} & \multicolumn{1}{c|}{-9.71} & -6.44 & -18.86 & \multicolumn{1}{c|}{0.14} & \multicolumn{1}{c|}{\textbf{0.17}} & \multicolumn{1}{c|}{0.06} & 0.14 \\ \hline
\multicolumn{1}{|c|}{\multirow{2}{*}{LibriSpeech-test}} & female & \multicolumn{1}{c|}{-10.04} & -5.00 & -20.13 & \multicolumn{1}{c|}{-0.05} & \multicolumn{1}{c|}{\textbf{-0.03}} & \multicolumn{1}{c|}{-0.07} & -0.07 \\
\multicolumn{1}{|c|}{} & male & \multicolumn{1}{c|}{-9.01} & -6.64 & -17.83 & \multicolumn{1}{c|}{-0.07} & \multicolumn{1}{c|}{\textbf{0.03}} & \multicolumn{1}{c|}{-0.01} & -0.07 \\ \hline
\multicolumn{1}{|c|}{\multirow{2}{*}{VCTK-test (dif)}} & female & \multicolumn{1}{c|}{-10.29} & -6.09 & -17.86 & \multicolumn{1}{c|}{0.60} & \multicolumn{1}{c|}{0.65} & \multicolumn{1}{c|}{-0.14} & \textbf{0.66} \\
\multicolumn{1}{|c|}{} & male & \multicolumn{1}{c|}{-11.69} & -8.64 & -17.95 & \multicolumn{1}{c|}{0.07} & \multicolumn{1}{c|}{\textbf{0.22}} & \multicolumn{1}{c|}{-0.10} & 0.10 \\ \hline
\multicolumn{1}{|c|}{\multirow{2}{*}{VCTK-test (com)}} & female & \multicolumn{1}{c|}{-9.31} & -5.10 & -20.39 & \multicolumn{1}{c|}{\textbf{0.21}} & \multicolumn{1}{c|}{0.12} & \multicolumn{1}{c|}{-0.07} & 0.19 \\
\multicolumn{1}{|c|}{} & male & \multicolumn{1}{c|}{-10.43} & -6.50 & -21.26 & \multicolumn{1}{c|}{-0.04} & \multicolumn{1}{c|}{\textbf{0.08}} & \multicolumn{1}{c|}{-0.03} & 0.00 \\ \hline
\multicolumn{2}{|c|}{Weighted average test} & \multicolumn{1}{c|}{-10.15} & -6.44 & -18.69 & \multicolumn{1}{c|}{0.11} & \multicolumn{1}{c|}{\textbf{0.18}} & \multicolumn{1}{c|}{-0.07} & 0.13 \\ \hline
\end{tabular}}

  \label{tab:gvd}
\end{table}

\begin{figure}[h]
    \includegraphics[width=0.95\linewidth]{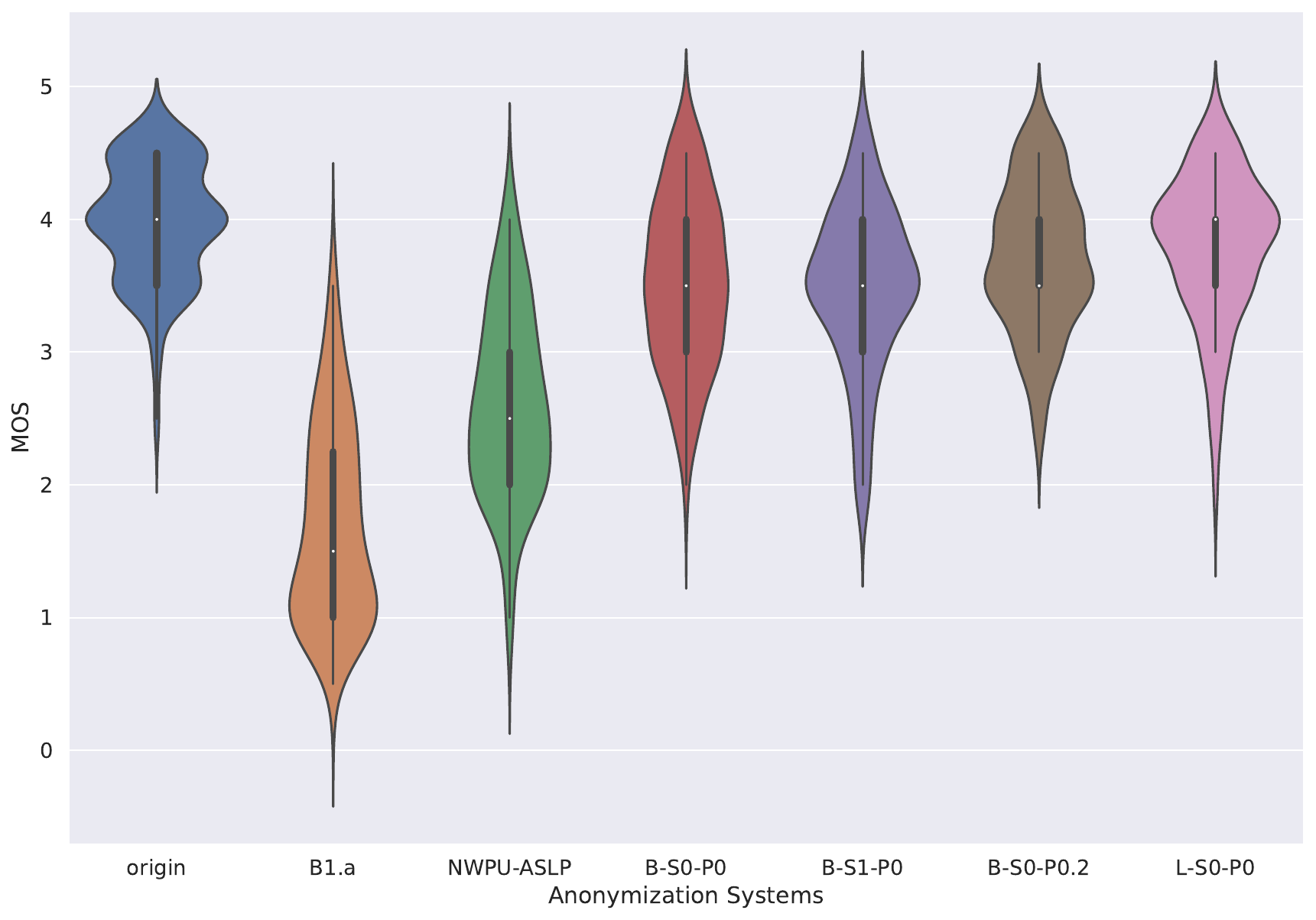}
    \caption{Mean Opinion Score (MOS) achieved on data processed by VPC baselines, NWPU-ASLP and our proposed system.}
    \label{fig:mos}
\end{figure}

\begin{table*}[htb]
  
  \renewcommand\arraystretch{1.5}
  \centering
  \footnotesize
    \caption{WER achieved by Whisper Large/U2++/TDNN-F on data processed by VPC baselines, NWPU-ASLP and our anonymization method.}
  \renewcommand{\tabcolsep}{0.11cm}
  \resizebox{0.90\textwidth}{!}{
  \begin{tabular}{|c|c|c|c|cccc|}
  \hline
  \multirow{2}{*}{\textbf{Dataset}} & \multicolumn{1}{c|}{\multirow{2}{*}{\textbf{Orig}}} & \textbf{VPC Baseline} & \textbf{VPC Top1}  & \multicolumn{4}{c|}{\textbf{SALT (Proposed)}}                                                                                              \\ \cline{3-8} 
                                    & \multicolumn{1}{c|}{}                               & \textbf{B1.a}         & \textbf{NWPU-ASLP} & \multicolumn{1}{c|}{\textbf{B-S0-P0}} & \multicolumn{1}{c|}{\textbf{B-S1-P0}} & \multicolumn{1}{c|}{\textbf{B-S0-P0.2}} & \textbf{L-S0-P0} \\ \hline
  LibriSpeech-dev                   & 5.12/3.80/3.82                                      & 13.00/5.14/4.34       & 7.00/4.94/\textbf{3.65}     & \multicolumn{1}{c|}{6.45/5.11/7.24}   & \multicolumn{1}{c|}{6.82/5.64/12.77}  & \multicolumn{1}{c|}{\textbf{6.37}/\textbf{4.69}/6.73}     & 6.68/4.73/6.73   \\ \hline
  VCTK-dev                          & 4.96/4.47/10.79                                     & 7.73/10.29/11.54      & 6.33/\textbf{6.51}/\textbf{7.62}    & \multicolumn{1}{c|}{7.12/10.68/20.19} & \multicolumn{1}{c|}{7.44/11.69/23.32} & \multicolumn{1}{c|}{\textbf{6.28}/8.62/18.79}    & 6.86/10.04/19.41 \\ \hline
  Average dev                       & 5.04/4.14/7.31                                      & 10.37/7.71/7.94       & 6.66/\textbf{5.72}/\textbf{5.63}     & \multicolumn{1}{c|}{6.79/7.89/13.71}  & \multicolumn{1}{c|}{7.13/8.67/18.04}  & \multicolumn{1}{c|}{\textbf{6.33}/6.66/12.76}    & 5.95/7.38/13.07  \\ \hline
  LibriSpeech-test                  & 6.02/3.76/4.15                                      & 13.16/4.92/4.75       & 6.77/4.81/\textbf{3.87}     & \multicolumn{1}{c|}{\textbf{5.71}/4.89/8.19}   & \multicolumn{1}{c|}{7.22/5.08/9.20}   & \multicolumn{1}{c|}{6.68/\textbf{4.58}/7.44}     & 6.21/4.66/6.11   \\ \hline
  VCTK-test                         & 4.41/3.32/12.82                                     & 6.54/7.53/11.82       & 5.49/\textbf{5.68}/\textbf{7.85}     & \multicolumn{1}{c|}{6.04/7.25/23.79}  & \multicolumn{1}{c|}{7.02/8.90/26.15}  & \multicolumn{1}{c|}{\textbf{5.45}/6.45/22.24}    & 5.68/7.05/21.64  \\ \hline
  Average test                      & 5.22/3.54/8.49                                      & 9.85/6.22/8.29        & 6.12/\textbf{5.25}/\textbf{5.86}     & \multicolumn{1}{c|}{\textbf{5.88}/6.07/15.99}  & \multicolumn{1}{c|}{7.12/6.99/17.67}  & \multicolumn{1}{c|}{6.07/5.51/14.84}    & 5.95/5.85/13.87  \\ \hline
  Overall                  & 5.13/3.84/7.90        & 10.11/6.96/8.12       & 6.39/\textbf{5.48}/\textbf{5.75}                  & \multicolumn{1}{c|}{6.33/6.98/14.85}  & \multicolumn{1}{c|}{7.12/7.83/17.86}  & \multicolumn{1}{c|}{\textbf{6.20}/6.08/13.80} & 5.95/6.62/13.47  \\ \hline
  \end{tabular}}

  \label{tab:wer}
  \vspace{-10pt}
\end{table*}

\section{Results}

We report the comparison results of our system and baselines in Sec. \ref{sec:comp}, followed by the PCA visualization of speaker embeddings under different scale factors in Sec. \ref{sec:viz}.

\subsection{Performance Comparison}
\label{sec:comp}
The EER results of the baseline and our proposed systems are given in Table \ref{tab:eer}. 
The results obtained from the proposed system reveal an average EER of 40\%, which stands as the highest EER compared to other baseline systems. The higher EER value demonstrates that the anonymized speech of our system can better anonymize personal speech data to protect user privacy.

Table \ref{tab:f0} and Table \ref{tab:gvd} show the intonation preservation and distinctive performance of different baseline systems and our proposed system. According to these results, our system achieves state-of-the-art performance in both $\mathrm{G_{VD}}$ and $\rho^\mathrm{F0}$ metrics. This indicates our system can keep the intonation and even enhance the speaker's distinctiveness of the original speech signal.

While achieving good results in the anonymization performance, our system can also preserve the quality and intelligibility of original speech. 
% According to the results in Table\ref{tab:wer} and Fig.\ref{fig:mos}, our system has a similar WER result compared to the baseline and the best MOS result among all tested systems, indicating that our system can well preserve the semantic information of original speech with good speech quality. 
According to the results in Fig.\ref{fig:mos}, our system has the best MOS result among all tested systems, indicating that our system can produce the most natural and comfortable anonymized speech.

We notice that in Table \ref{tab:wer}, the WER results differ across multiple ASR models. We believe the speech accent is the primary reason since our system utilizes the joint feature of context and speaker information and can preserve the accent information, while other baseline systems lost the most of accent information during the context-speaker decoupling. We can find the WER results of our model in Whisper and U2++ are better than the VPC baseline system, due to both Whisper and U2++ models are trained in large-scale datasets and are more robust than the VPC ASR model in accent data recognition.

% Due to the training set scale of three ASR models, the accent robustness varies across these models, which causes the WER differences.

We also notice that using features extracted by WavLM-Large leads to higher quality and intelligibility metrics, which means a better encoder leads to better performance.

We further verify the diversity-quality trade-off aforementioned in Sec.\ref{sec:matcher}, according to previous results, when we increase the scale factor, we observe that the diversity and anonymization metrics such as $\mathrm{G_{VD}}$ and EER become better while quality metrics such as $\rho^\mathrm{F0}$, WER and MOS becomes worse, and when we increase the preservation factor, we observe the opposite results. These results show our system can tradeoff between diversity and quality by adjusting the $s$ and $p$ factors.

\begin{figure}[h]
\centering
\includegraphics[width=0.60\linewidth]{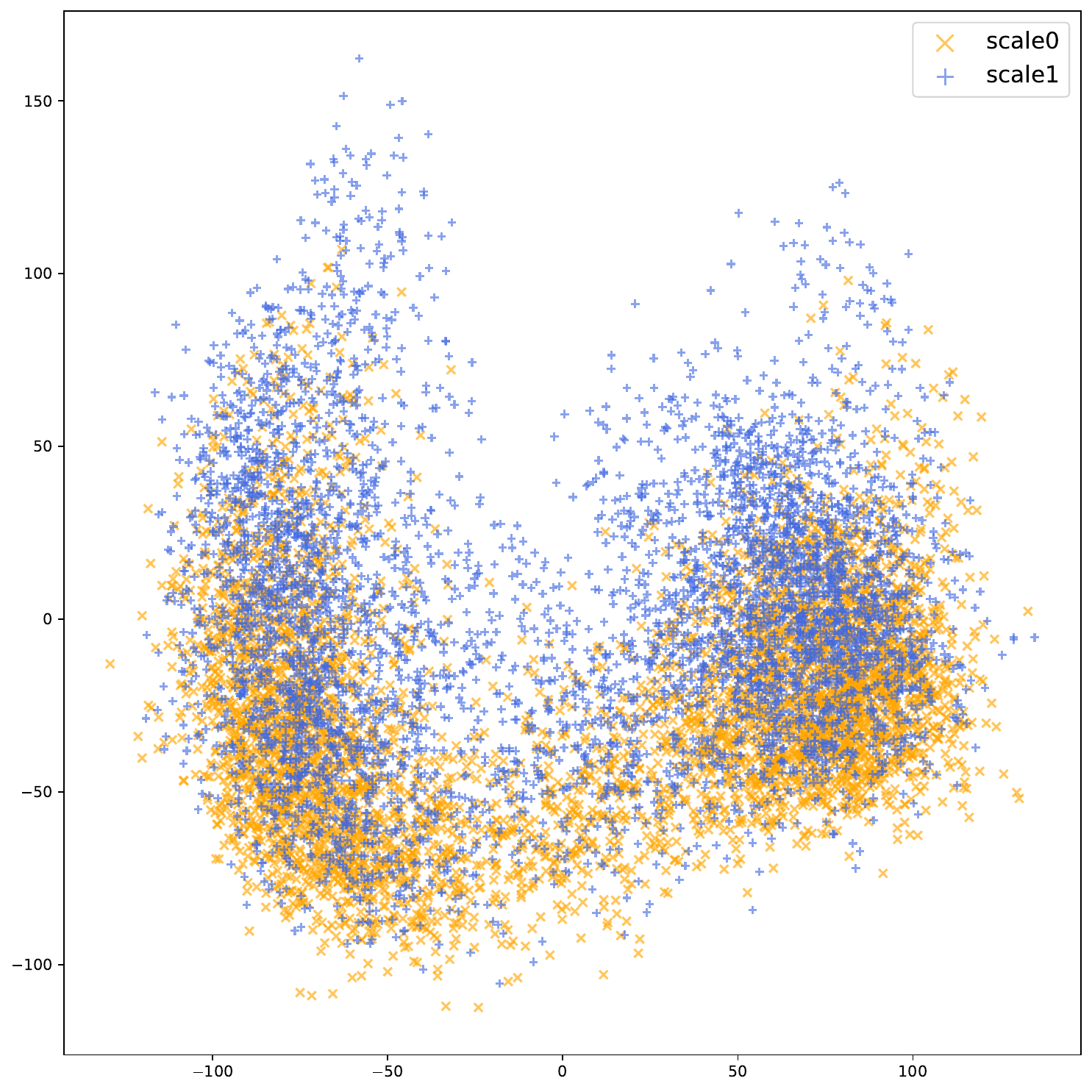}
\caption{Speaker visualization of anonymized speeches, where the orange dots indicate the scale factor is 1, while blue dots indicate the speaker features are not extrapolated.}
\label{fig:pca}
\end{figure}

\subsection{Visualization}
\label{sec:viz}

To better show the effectiveness of our proposed extrapolation anonymization method, we randomly anonymize 5000 speeches from librispeech-train-clean-360, and then we employ PCA projection of the speaker embedding vectors which is extracted from anonymized speeches of B-S0-P0 and B-S1-P0. We use ECAPA-TDNN~\cite{DBLP:conf/interspeech/DesplanquesTD20} from Speechbrain~\cite{DBLP:journals/corr/abs-2106-04624}\footnote{\url{https://github.com/speechbrain/speechbrain}}, an ASV model pre-trained on VoxCeleb1+2, as our extractor of speaker embedding vectors. Fig.\ref{fig:pca} shows the visualization of PCA projected speaker embedding. We observe that the speaker embedding diversity of B-S1-P0 is better than B-S0-P0. This shows the extrapolation of speaker features can lead to a wider distribution of pseudo speakers.

\section{Conclusion}

In this paper, we proposed \modelname{}, a speech anonymization system based on latent space transformation. The proposed system uses a WavLM encoder, kNN feature matching, and random weighted averaged latent feature to obtain the anonymized feature. Then a modified vocoder is used to transform the latent feature into an anonymized speech signal. In addition, we have proposed feature extrapolation and feature reservation, which allow the proposed model to trade-off between the diversity and quality of the anonymized speech. Following the same evaluation setup as VPC 2022, our system achieves the best EER, $\rho^\mathrm{F0}$ and $\mathrm{G_{VD}}$ of 40\%, 0.79, and 0.12, respectively. In particular, our system achieves a positive $\mathrm{G_{VD}}$ metric, which is also state-of-the-art. The results show that our system can transform a speaker's speech into a distinguishable anonymous speech while preserving speech quality and intelligibility.

\bibliographystyle{IEEEbib}
\bibliography{strings,refs}

\end{document}